\documentclass[11pt]{article}
\usepackage{amssymb}

\usepackage{graphicx}
\usepackage{amsmath}

\newcommand{\PP}{{\mathbb{P}}}
\newcommand{\Z}{\mathbb{Z}}
\newcommand{\C}{\mathbb{C}}
\newcommand{\R}{\mathbb{R}}
\newcommand{\Q}{\mathbb{Q}}
\newcommand{\M}{\overline{M}}
\newcommand{\Aut}{\operatorname{Aut}}
\newcommand{\la}{\lambda}

\begin{document}

\title{Open-String Gromov-Witten Invariants:\\
Calculations and a Mirror ``Theorem''}
\author{Tom Graber and Eric Zaslow}
\maketitle

\begin{abstract}
We propose localization techniques
for computing Gromov-Witten invariants
of maps from Riemann surfaces with
boundaries into a Calabi-Yau,
with the boundaries mapped to a Lagrangian submanifold.
The computations can be expressed in terms of
Gromov-Witten invariants of one-pointed maps.
In genus zero,
an equivariant version of the mirror theorem
allows us to write down a hypergeometric
series, which together with a mirror map allows
one to compute the invariants to all orders,
similar to the closed string model or the physics
approach via mirror symmetry.
In the noncompact example where the
Calabi-Yau is $K_{\PP^2},$
our results agree with physics predictions
at genus zero obtained using
mirror symmetry for open strings.
At higher genera, our results satisfy strong
integrality checks conjectured from physics. 
\end{abstract}


\section{Introduction}

\subsection{The Physics}

Mirror symmetry is famous
for being able to predict Gromov-Witten
invariants of Calabi-Yau manifolds.  The basic
conjecture is that there is a duality
between string theories on mirror Calabi-Yau
manifolds.  As a consequence,
the topological field theory defined
from the A-twist of one Calabi-Yau manifold is equal
to the topological B-twist of the mirror.  Both twists
can be performed on Calabi-Yau target manifolds.
From a practical point of view, in order to obtain enumerative
predictions, one needs to know the theory on the B-model
(in this case, defined through classical period integrals)
as well as an identification of the parameter spaces
for both theories -- the ``mirror map.''  To extract
integer-valued invariants, one needs an all-genus
``multiple-cover'' formula.
The technology for finding mirror manifolds \cite{Batyrev} and
performing calculations of the mirror map \cite{HKTY}
is well established.  The multiple-cover formula was
found at genus zero in \cite{CDGP} and generalized to
all genera in \cite{GV}.  At higher genus, the BCOV equations
\cite{BCOV} can be used (up to an ambiguity) to calculate Gromov-Witten
invariants.  Mathematical verification of these predictions
has only be performed in the noncompact setting
\cite{Hosono1} \cite{KZ}.
The integer invariants
are known as Gopakumar-Vafa invariants, and still await
a mathematical definition.\footnote{See \cite{BP}, \cite{KKV},
and \cite{Hosono2} for progress in this direction.}

String theory
in a spacetime-filling brane is described by maps from
Riemann surfaces in which the boundaries are mapped to
submanifolds in the Calabi-Yau.  Other data such as gauge
fields on the submanifolds may be present.
Not all submanifolds preserve the supersymmetries necessary
to perform the topological twisting.
Lagrangian submanifolds preserve the A-twist,
while holomorphic submanifolds preserve the B-twist.
The mirror symmetry duality implies that to each A-brane
there corresponds a mirror B-brane, so that the
resulting string theories are still equivalent.  As a consequence,
the resulting topological field theories must also be the
same.  On the B-model side, this resulting theory
is known \cite{BCOV} \cite{Witten} 
to be given by a holomorphic version of Chern-Simons theory.

The insight of Aganagic and Vafa
\cite{AV} was an identification of
special pairs of mirror branes in the setting where one
Calabi-Yau was a noncompact toric variety,
as well as dimensional
reduction of the holomorphic Chern-Simons theory.
Those authors chose a Lagrangian A-brane whose corresponding
B-brane was a holomorphic curve inside the mirror.
(Such pairs were also described in \cite{Hori}.)
Though holomorphic Chern-Simons theory is defined for complex
three-folds, the reduction to complex curves 
can be performed, and results in a computable integral of a meromorphic
two-form.  Aganagic and Vafa
then found a mirror map
identifying brane moduli, so that the expansion of a superpotential
in the mirror coordinates encoded the (conjectural) Gromov-Witten
invariants involving holomorphic
maps from Riemann surfaces with boundaries
to the noncompact Calabi-Yau, where
the boundaries get mapped
to the relevant Lagrangian submanifold.
What resulted was a prediction of these would-be
open-string Gromov-Witten
invariants at genus zero -- i.e., for maps from disks.
Such invariants have not yet been defined rigorously, from a
mathematical point of view.\footnote{Insofar as these
invariants are
related to Fukaya-Floer theory, much of the work towards making
a rigorous mathematical theory of them has been carried out in \cite{FOOO}.}

The multiple-cover formula for open-string
Gromov-Witten invariants
was found in \cite{OV}.  This has been generalized to
Riemann surfaces with more than one boundary component
in \cite{LMV}.
These formulas have been verified through
localization calculations in \cite{LS} and \cite{KL}.
Recently, Aganagic, Klemm and Vafa \cite{AKV}
expanded this mirror
procedure (``B-model'')
in a number of new examples, and were able to make
integer predictions of the invariants.  
Mari\~no and Vafa \cite{MV} have done the same for
higher-genus, multiple-boundary invariants
using Chern-Simons duals.
In this note we will
perform localization calculations to provide an
explicit ``A-model''
verification of these predictions,
and use an equivariant mirror theorem \cite{E} to compute.
The results we obtain match those
authors' perfectly, including a dependence on an
additional $\Z$-valued parameter.

In addition, we are able to extend our calculations to higher
genus Riemann surfaces with boundary, and to multiple
boundary components.  No physical predictions
have been made of these numbers, though there are
strong integrality properties predicted through the
work of Ooguri and Vafa \cite{OV}
and of Labastida, Mari\~no and Vafa \cite{LMV}.  These
requirements have been met in all compuatations checked.

\subsection{The Math}

In \cite{Kont},
Kontsevich defines a moduli space of stable maps,
which allows one to compute Gromov-Witten invariants
in many (toric) examples through
localization techniques.  The basic idea
of these calculations is straightforward.  While the
moduli space of maps is quite complicated, the
subspace which is fixed under the natural
torus action is simple.  By applying a localization
theorem we can turn integrals over the entire
space of maps into integrals over the fixed
locus which we can compute.

Kontsevich used this method to 
compute genus zero Gromov-Witten
invariants of a homogeneous space.  In this
case, the space of maps is smooth, and the
Bott residue theorem suffices to reduce the
computation of the invariants to integrals over
$\M_{0,n}$.

For higher genus invariants, or nonconvex target
spaces, a virtual localization theorem is needed,
since the Gromov-Witten invariants 
are defined to be integrals against a
virtual fundamental class.
In \cite{GP}, such a theorem is proven, provided
the target space is a smooth algebraic variety with
an algebraic $\C^*$ action.

The current setting, however, is non-algebraic.
Moreover, there is not (yet) a mathematical formulation
of the Gromov-Witten invariants associated to maps
from Riemann surfaces with boundary.
One then hopes to define a virtual fundamental
class and prove a localization theorem for
this class in 
a symplectic setting.
We do {\bf not} prove such a theorem in this paper.
Instead, we perform a calculation in the spirit of
localization, and find agreement with the
predictions of \cite{AKV}.


We now briefly describe the calculation.  The
Calabi-Yau we consider is $K_{\PP^2}$, the
total space of the canonical bundle of $\PP^2$,
and the Lagrangian submanifold $L$ is constructed from the
co-normal bundle over a straight line in the toric
polytope (the image of the
moment map used in the construction of $K_{\PP^2}$).  It intersects
the $\PP^2$ in an $S^1$ which we label $S^1_L$.
This Lagrangian is preserved under a real torus, which
acts on the moduli space of maps.
The fixed points correspond to
source curves which
map equivariantly to invariant curves in the image.
The component of such
a source curve containing the
(single) connected boundary component must map to
a disk in $\PP^2$ ending on $S^1_L,$ and is determined by the
sign of the winding of the loop around $S^1_L.$ 
In this way, the fixed-locus components should be described by
the graphs that Kontsevich introduced, with a single ``leg''
attached corresponding to the boundary disk.
To each graph $\Gamma$
is associated a moduli space $M_\Gamma$
which is a product over vertices $v$ of
${\mathcal M}_{genus(v),valence(v)},$
i.e. the moduli of the contracted components.
The integrals are performed using Faber's algorithm.

The alert reader may note that the leg is specified
by an attachment location and a winding parameter.
Indeed, we define a winding-dependent integral involving
stable maps with one marked point, which, together
with a winding-dependent multiplicative constant,
yields the same result.  This observation, together
with an equivariant mirror theorem, can be
used to write a hypergeometric series and mirror map
which computes all genus-zero open-string Gromov-Witten
invariants.

\vskip0.2in
In the next section, we describe more carefully the
geometric setting
which will serve
as the main example of the localization
formulas obtained in Section \ref{weights}.
Genus-zero invariants are computed with the equivariant
mirror theorem in Section \ref{mirsec}, and a comparison with
the physics open mirror symmetry procedure is made.
We conclude with some example calculations involving
integer invariants (BPS numbers),
and some comments about future directions.

\section{Defining Terms}
\label{sec:def}
Define the ``symplectic quotient''
\begin{equation}
\label{defx}
M \equiv \lbrace |z_1|^2 + |z_2|^2 + |z_3|^2 - 3|z_4|^2 = r
\rbrace /S^1,
\end{equation}
where $r>0$ and $S^1$ acts by $\theta: (z_1,...,z_4)\mapsto
(e^{i\theta} z_1,e^{i\theta} z_2,e^{i\theta} z_3,
e^{-3i\theta} z_4),$ i.e. with weights $(1,1,1,-3).$
Then $M\cong K_{\PP^2},$ the total space of the canonical
bundle of $\PP^2$.
Note that $z_4$ parametrizes the fiber, and that $z_1,$ $z_2,$
and $z_3$ cannot be simultaneously zero.

In the patch $U \equiv \{z_3 \neq 0\},$ we can parametrize $M$ with three coordinates
$z,$ $n,$ and $f$ defined by
$$z = a z_1/z_3, n = z_2/z_3, f = z_4 z_3^3,$$
where the constant $a = \sqrt{(r+c)/(r-c)}$ is included
for later convenience
($c<r$ is defined in Eq.\!\!\!(\ref{lageq})).
Note that $z,n,f$ determine $z_1, ..., z_4$ via Eq.\!\!\!
(\ref{defx}), up to the quotiented $S^1$ action.

There is a residual $(S^1)^3$ action on $M.$  We will exploit this
later to perform localization calculations.  In fact, it is often
enough to consider a generic subgroup $S^1 \subset (S^1)^3,$
which we define by weights $\lambda_z, \lambda_n, \lambda_f.$
Explicitly, $\theta: (z,n,f) \mapsto (e^{i\lambda_z \theta}z,
e^{i\lambda_n \theta}n,e^{i\lambda_f \theta}f).$

The symplectic form $\omega
= \frac{1}{2} \sum_{i=1}^{4} d|z_i|^2 \wedge d\theta_i$
descends to $M,$  and we define the Lagrangian
submanifold $L$ by
\begin{eqnarray}
\label{lageq}
|z_1|^2 - |z_3|^2 = c, \cr
|z_2|^2 - |z_4|^2 = 0, \cr
{\rm Arg}(z_1 z_2 z_3 z_4) = 0, \cr
\end{eqnarray}
where $c<r.$
Note that these equations make sense on $M.$
The $S^1$ action doesn't affect the absolute values,
but in order to preserve $L,$ we must require
that
\begin{equation}
\label{weightcond}
\lambda_z + \lambda_n + \lambda_f = 0.
\end{equation}
In terms of weights $\lambda_i$ for the $z_i,$
we can write $\lambda_z = \lambda_1 - \lambda_3,$
$\lambda_n = \lambda_2 - \lambda_3,$
and $\lambda_f = - (\lambda_n + \lambda_z) =
2\lambda_3 - \lambda_1 - \lambda_2.$
Note that $L$ is obtained by a co-normal construction
of a linear subspace of the image of the moment
maps $|z_i|^2/2.$

Define $\PP^2$ by $z_4 = 0$ and note that $L\cap \PP^2$
is a circle $S^1_L$ defined by
$f = n = 0,$ $|z| = 1.$ 
Noting that $L$ has the topology of $S^1 \times \R^2,$
we coordinatize $L$ with a real angular variable $\theta$
and a complex variable $x$ (this is {\bf not} meant
to suggest that $L$
has a complex structure).
Let us define $\varphi: L\hookrightarrow M$ in the
coordinates $(z_1,...,z_4)$ by
$$\varphi(\theta,x) = (\sqrt{|x|^2 + \frac{r+c}{2}}\; e^{i\theta}
e^{i\theta_3},
xe^{i\theta_3},\sqrt{|x|^2 + \frac{r-c}{2}}\; e^{i\theta_3},
\overline{x}e^{-i\theta}e^{-3i\theta_3}).$$
Note that $\theta_3$ parametrizes the $S^1$ ambiguity,
whereas the values
of $z,n,f$ are well-defined.
Note also that $S^1_L$ is the image
of $S^1\times (0,0)\subset S^1\times \R^2 = L$

One calculates that {\em along} $S^1_L,$
\begin{eqnarray}
\label{push}
\varphi_*(\frac{\partial}{\partial \theta}) =
\frac{\partial}{\partial \theta_z} =
iz\frac{\partial}{\partial z} - i\overline{z}\frac{\partial}{\partial
\overline{z}}\cr
\varphi_*(\frac{\partial}{\partial x}) =
\frac{\partial}{\partial n} + e^{-i\theta_z}
\frac{\partial}{\partial \overline{f}}
\end{eqnarray}

We will be interested in
holomorphic maps from a disk into $M,$ such that
the boundary of the disk is mapped into $L.$  The infinitesimal
deformation of such a map is a holomorphic section of the
pull-back of $TM$ which lives in the pull back of $TL \subset TM$
along the boundary.  Therefore, we now describe how $TL$ sits
inside $TM.$  In fact, we will only need to describe $TL|_{S^1_L}
\subset TM|_{S^1_L}$ since the boundary of the disk will land inside
$S^1_L$ via a holomorphic map.  As $S^1_L$ is contained in our patch
$U,$ we can identify $TM$ with the trivial bundle
$\C^3$ generated by
$\frac{\partial}{\partial z},$
$\frac{\partial}{\partial n},$
$\frac{\partial}{\partial f}.$

The holomorphic map of most
interest to us will be
\begin{equation}
\label{psieq}
\psi_w: D\rightarrow M,
\end{equation}
where $D$ is the unit disk $\{|u|\leq 1\}$ and $\psi_w(u)=
(z = u^w,n = 0,f = 0).$ 
Note that $\psi_w: \partial D \rightarrow S^1_L$
has winding number $w$.  (A similar map based at an
opposite pole of an invariant $\PP^1$ can be defined
when $w<0.$)
When we pull back $TM$ to $D$ via $\psi_w,$ the
totally real sub-bundle
$TL\subset TM$ can be described by the real span of a basis
defined by a unitary matrix acting on the frame
$\frac{\partial}{\partial z},$
$\frac{\partial}{\partial n},$
$\frac{\partial}{\partial f}.$
Let $\theta$ parametrize $\partial D.$
Then over the point $u = e^{i\theta}\in \partial D,$
we see from Eq.\!\!\!(\ref{push}) that
$TL$ is defined by the matrix
\begin{equation}
\label{Amatrix}
A = \frac{1}{\sqrt{2}}\begin{pmatrix}
i\sqrt{2}e^{iw\theta}&0&0\\
0&1&i\\
0&e^{-iw\theta}&-ie^{-iw\theta}\\
\end{pmatrix}.
\end{equation}
Note that this matrix is unitary, and its determinant is constant --
a consequence of the special Lagrangian condition
that the phase of the holomorphic three-form is constant along $L.$

\section{Localization Calculation}
\label{weights}

The authors of both \cite{KL} and \cite{LS} find an
algebraic moduli space in which the space of stable
maps from Riemann surfaces with boundary embeds.
One hopes to be able to approach the general
non-algebraic problem via a virtual localization
formula (not proven) for this space.
In the following, we assume some such formula is
valid (see Eq.\!\!\! (\ref{virloc})).
In fact, our calculations will ultimately be
torus weight-{\em dependent}, producing a discrete
dependence of the answers which matches
physical predictions.
The origins of this dependence are unclear to us.
However, the actual calculation is straightforward.
As in the closed string case, while the global geometry
of the moduli space of maps is complicated, the
geometry of the fixed locus is simple.  Indeed,
the fixed loci are identical to those that arise
in space of maps from closed Riemann surfaces.
(We exploit this observation in Section \ref{mirsec}.)
Moreover, the deformation theory that governs
the contribution of such a fixed locus to
the integral is essentially the same
as in the closed case.  As a result, even
in the absence of a general theory to justify
such a computation, we have a natural method
for finding these invariants.


Let us describe the localization calculation.
As in \cite{Kont}, the fixed-point locus will be
described by decorated
graphs which relate: the invariant $\PP^1$'s where
genus-zero components of the source curves are
mapped; the fixed points in $\PP^2$ where
contracted components are mapped; and the
disks sitting inside invariant $\PP^1$'s where
disk components are mapped.  We therefore define
a decorated graph
$\Gamma = ({\mathcal V},{\mathcal E},
{\mathcal L})$
as triple of
vertices ${\mathcal V}$, edges $\mathcal E,$ and ``legs'' 
$\mathcal L$ defined as
follows.
Each vertex $v\in\mathcal V$ carries a label $i(v)$
of a torus fixed point among
$p_1,...,p_3$
in $\PP^2$ (here $p_i$ is the point in $\PP^2$
with $z_i$ as the only non-zero coordinate),
as well as the genus $g(v)$ of a contracted
component $C_v$ of the source curve. 
Each edge $e\in \mathcal E$ carries
a positive degree label $d_e$
of the map from the edge component $C_e,$
and an unordered pair $(i(e),j(e))$ of vertices that
the edge joins.  A ``leg'' $l\in \mathcal L$
will represent a disk source component $C_l$ with
boundary landing in $L\cap \PP^2,$ i.e., $S^1_L$
in the example in this paper.  Each leg is labelled
by a winding $w(l)$ describing the degree of the map
from $\partial C_l$ to $S^1_L.$  In addition,
we may wish to refer to $i(l),$ a label of the fixed point
that the center of the disk $C_l$ is mapped to,
and $j(l),$ the fixed point at the other pole of the
invariant $\PP^1\subset \PP^2$ containing the image
of $C_l.$
In
the example of this paper, positive winding will demand
that $i(l) = p_3$ while negative winding demands $i(l) = p_1.$ 
The valence of a vertex, $val(v),$ will
mean the number of
edges and legs which meet it.
Also, as we consider
Riemann surfaces with connected boundary, the leg
set will consist of exactly one element.  Nevertheless,
these methods extend to the multi-component
boundary case as well.
Stability, the total genus, and the relative homology
class of the image curve
(i.e., degree and winding) place the usual restrictions on
graphs (cf. \cite{Kont}).

Each graph
$\Gamma$ defines a component of the fixed-point locus
$M_\Gamma = \prod_{v}\overline{M}_{g(v),val(v)}.$
A virtual localization formula
\`a la \cite{GP} for open-string Gromov Witten
invariants $K^g_{d,w}$
would take the form
\begin{equation}
\label{virloc}
K^g_{d,w} = \int_{\overline{\mathcal M}^{vir}} 1
= \sum_{\Gamma}\frac{1}{|{\bf A}_\Gamma |}
\int_{M_{\Gamma}}\frac{1}{e(N_\Gamma^{vir})},
\end{equation}
where $e(N_\Gamma^{vir})$
represents the Euler class of the
virtual normal bundle of $M_\Gamma$ inside ${\mathcal M},$
$i$ is the inclusion map from the fixed-point locus,
and
${\bf A}_{\Gamma}$
is an automorphism group which is defined as
the group $\Z/w\Z \times {\bf A}_{\Gamma '}$ 
where $\Gamma '$ is the graph with the leg deleted
and ${\bf A_{\Gamma '}}$ is the automorphism group
defined in \cite{GP}.  Concretely, the order of
this group, which is all that is relevant here, is
$w|\Aut (\Gamma)| \times \prod_{{\rm edges}} d_e$.

What remains is to calculate the weights of the
torus action on the various vector spaces and bundles
over the fixed-point loci.
The basic exact sequence
which describes the ``obstruction bundle''
calculation is
$$0\rightarrow
Aut(C)\rightarrow H^0(\psi^*TM)\rightarrow T{\cal M}\rightarrow
Def(C)\rightarrow H^1(\psi^*TM)\rightarrow Obs\rightarrow 0.$$
We interpret this sequence as a sequence of bundles
fibered over the moduli space.
In this sequence, $Aut(C)$ and $Def(C)$
are ${\rm Ext}^k(\Omega_C(D),{\cal O}_C),$
$k = 0,1,$ respectively, where $D$ is the
divisor of marked points (the nodes, in our example). 
When $C$ is smooth these spaces are familiar as
$H^0(TC)$ and $H^1(TC).$  The notation $H^k(\psi^*TM),$
$k=0,1,$ is short for $H^k(C,\partial C;\psi^*TM,\psi^*TL)$
over a point $\psi:C\rightarrow M$ in
the moduli space, ${\mathcal M}.$
For example, when $C$ is a smooth disk, $H^0(\psi^*TM)$
represents global holomorphic tangent vector-valued
sections over the disk, which lie in the real
sub-bundle $TL$ along the boundary $\partial C\cong S^1$
This sequence tells us that
\begin{equation}
\label{numden}
\frac{Obs}{T{\cal M}} =
\frac{Aut(C)}{Def(C)}\cdot \frac{H^1(TM)}
{H^0(TM)}.
\end{equation}
To compute the right-hand side we use the normalization
for $C$ in terms of the irreducible components $C_i$
(here $C_i$ represents $C_v,$
$C_e,$ and $C_l$):
\begin{equation}
\label{normseq}0\rightarrow {\cal O}_C\rightarrow
\bigoplus_{i}{\cal O}_{C_i}\rightarrow
\bigoplus_n{TM|_{\psi(x_n)}}
\rightarrow 0,
\end{equation}
where $n$ runs over the nodes $x_n$ lying at the
intersection of the components.  In what follows
we will use
the related notation of ``flags'' $F,$ with the
convention that a flag can lie at the intersection of
$C_v$ and either an edge {\em or} a leg.

We get all the information we need when we tensor
Eq.\!\! (\ref{normseq}) by $(\psi^*TM,\psi^*TL):$

\begin{equation}
\label{honehzero}
\frac{H^1(TM)}{H^0(TM)} =
\frac{\bigoplus_i H^1(C_i,\psi^*TM)
\bigoplus_n T_{\psi(x_n)}M}{\bigoplus_i H^0(C_i,\psi^*TM)}.
\end{equation}
The calculation has therefore been reduced to computations
of $Aut(C),$ $Def(C),$ and
sections of bundles
over the irreducible components of the fixed-loci
source curves.  Most of this is well-trodden material,
and involves keeping track of the torus weights
of various line bundles over $\PP^1$ or twisted
Hodge and cotangent line
bundles over contracted components.
In particular, the $Aut(C),$ $Def(C),$
and denominator terms
in
Eq.\!\!\! (\ref{numden})
{\em which do not involve legs}
have been
discussed in \cite{GP}.
The other numerator terms not involving legs have
been discussed in \cite{KZ}.
The torus weights of legs
have been discussed in \cite{KL};
we now provide an alternate derivation.

Let $C_l$ be a leg component with $\psi$ the map
of Eq. (\ref{psieq}).  We are interested in the vector
spaces $H^k(C_l,\partial C_l; \psi^*TM,\psi^*TL),$ $k=0,1.$
Let us search for $H^0,$ the holomorphic sections.  (A
similar analysis can be applied to $H^1$.)
Since $C_l$ is mapped entirely
within the open set $U$ coordinatized by
$z, n, f,$ the bundle $TM$ is trivial and we can
describe holomorphic sections by a triplet
$h = (h_1, h_2, h_3)$ of holomorphic functions of $u.$
The boundary condition described by the 
matrix $A$ in Eq. (\ref{Amatrix})
states that on the boundary,
$$h(e^{i\theta}) = A\cdot a,$$ 
where $a$ is a triplet of {\em real} functions of $\theta.$
Since $A$ is block diagonal, we can look at the blocks
separately.  The northwestern $1\times 1$ block states
$h_1(e^{i\theta}) = a_1 i e^{iw\theta},$
so $-ie^{-iw\theta} h_1$ is a real function.
Since $h_1$ is holomorphic, we can expand
$h_1(e^{i\theta}) =
\sum_{k\geq 0}h_{1,k}e^{ik\theta},$
and we then have the equation
$$-ie^{-iw\theta}\sum_{k=0}^{\infty}h_{1,k}e^{ik\theta}=
ie^{iw\theta}\sum_{k=0}^{\infty}
\overline{h_{1,k}}e^{-ik\theta},$$
from which we derive $h_{1,k} = -\overline{h_{1,2w-k}},$
$k = 0,...,2w,$ with all others vanishing.
The lower block can be treated similarly; one
finds that there are no contributions to
$H^0,$ while only the lower block contributes
to $H^1.$
As a result, we find that $H^0$ is a
real $(2w+1)$-dimensional vector space, and recalling
that $u$ has weight $\lambda_z/w,$ we see that the
action of $S^1$ on $H^0$ decomposes into representations
of the type $\begin{pmatrix}0&m\\ -m&0\end{pmatrix},$
$m = 1,...,w,$ and one trivial representation
(whose zero weight will cancel against an automorphism).
Up to an overall sign, then, we have found
that the equivariant Chern class
of the vector space
$H^0$ is $\prod_{m=1}^{w}(m/w)(\lambda_3-
\lambda_1),$
where we have used $\lambda_z = \lambda_1-\lambda_3.$
This combination appears in the denominator of the
first line of the equation below.
As discussed in \cite{FOOO}, the orientation
on the vector space $H^k(C_l,\psi^*TM)$
is a subtle issue affecting an overall sign
in our calculation.  Without a rigorous moduli
space to work with, we are unable to determine this
sign directly, so we choose
a sign convention which leads to agreement with
physics and integrality.

Putting these contributions
together, one arrives at a formula
for the integrand of $\int_{M_\Gamma}$ in Eq.\!\!\!
(\ref{virloc}).
In the following, we have written $k(l)$ for the
unique fixed point in $\PP^2$ not equal $i(l)$
or $j(l).$  Also, $\omega_F \equiv
(\lambda_{i(F)}-\lambda_{j(F)})/d,$ where
$d = d_e$ for an edge flag and $d = w(l)$
for a leg flag.  $e_F$ is the first Chern class
of the cotangent line bundle over $M_\Gamma$
associated to the point
$x_F \in C_v.$  We also write the top Chern
class of a twisted dual Hodge bundle as
$P_g(\lambda,E^*) = \sum_{r=0}^{g}\lambda^r c_{g-r}(E^*),$
where $E$ has fibers $H^0(K_{C_v}).$
In total, we have the following formula,
in which the $\frac{H^1}{H^0}$
leg contributions are on the
first line (for simplicity we have written the
formula for general {\sl positive} windings only),
the non-leg $H^1(K_{\PP^2})$ contributions
on the second line, and the $e(N^{vir}_\Gamma)^{-1}$
contributions on the following lines
(we have performed all genus-zero moduli space
integrations, using Kontsevich's formula in \cite{Kont}):
\begin{eqnarray*}
\label{virnorm}
\frac{i^*\phi}{e(N^{vir})} &=&
\prod_{l}
\frac{\prod_{m=1}^{w(l)-1}
\left[(-m/w(l))(\lambda_{i(l)}-\lambda_{j(l)})
-(\lambda_{k(l)}-\lambda_{j(l)})\right]}
{\prod_{m=1}^{w(l)}(-m/w(l))(\lambda_{i(l)}
-\lambda_{j(l)})}
\cr
&\phantom{=}&
\times
\prod_{v}\Lambda_{i(v)}^{val(v)-1}P_{g(v)}(\Lambda_{i(v)},E^*)
\prod_{e}\left[\prod_{m=1}^{3d_e - 1}\Lambda_{i(e)} +
\frac{m}{d_e}
(\lambda_{i(e)} -\lambda_{j(e)})\right]\\
&\phantom{=}&
\times
\prod_{e}\frac{(-1)^{d_e} d_e^{2 d_e}}{(d_e!)^2 
(\lambda_{i(e)}-\lambda_{j(e)})^{2 d_e}}
\prod_{{a+b=d_e\atop a,b\ge 0}\atop k\neq i(e),j(e)}
{\frac{1}{\frac{a}{d_e} \lambda_{i(e)}+\frac{b}{d_e}
\lambda_{j(e)}-\lambda_k}}\\
&\phantom{=}&\times \prod_{v} \prod_{j\neq i(v)} \, 
(\lambda_{i(v)}-\lambda_j)^{val(v)-1}\\
&\phantom{=}&
\times
\begin{cases}
\displaystyle
\prod_{v}
\left[\left(\sum_{F} w_F^{-1}\right)^{val(v)-3}\prod_{F\ni v}
w_F^{-1}\right]
\qquad\qquad\quad\;\;\, {\rm if }\; g(v)=0, \cr
\displaystyle
\prod_{v}
\prod_{j\neq i(v)} \,
P_{g(v)}(\lambda_{i(v)}-\lambda_j,E^*) 
\prod_{F\ni v} \frac{1}{w_F-e_F}
\qquad {\rm if}\; g(v)\ge 1,
\end{cases}
\end{eqnarray*}
where $\Lambda_i
\equiv\lambda_1 + \lambda_2 + \lambda_3 - 3 \lambda_i.$
As a small check, one can see that the $\Lambda_i^{-1}$
term of the second line and the $(\lambda_{i(v)}-
\lambda_{j(v)})^{-1}$ terms of the fourth line
come from the node contributions to $H^1(\psi^*TM)$
in the numerator of Eq.\!\!\! (\ref{honehzero}).

These formulas can be generalized for Lagrangians
in other noncompact, toric Calabi-Yau geometries
as well.

\section{Mirror Symmetry for Open String Invariants}
\label{mirsec}

The techniques of the previous section apply in all
genera, but in genus zero more can be said.
Here we show how a simple observation, together
with the equivariant mirror theorem,
leads to a kind of ``proof'' of mirror symmetry
in genus zero.

We return to the geometry of Section \ref{sec:def}.
We will consider open-string Gromov-Witten invariants
for stable maps with a single boundary component,
and with a positive winding, so that the leg at the fixed
locus is forced to attach at the ``north pole,'' $p$
($z=n=f=0$).
As we now show,
an inspection of the contribution of the legs
in Eq. \!\!(\ref{honehzero}) shows that the calculation
of the Gromov-Witten invariant differs only by a
winding-dependent leg
factor from a one-pointed stable map calculation,
where the marked point is forced to map to the north
pole.  The reason is simple.  The fixed loci are
described by the same data:  knowing where the marked
point is on the source curve tells you where to attach a
leg with winding $w.$  An overall factor will account
for the contribution of the leg to
Eq. \!\!(\ref{honehzero}).  What remains is the fact
that the tangent to the source curve
at the attached point affects
$Def(C).$  We therefore write down a one-pointed
Gromov-Witten invariant which is tailor-made
to contribute the appropriate factor in $Def(C)$
at the marked
point.

Let $K_{d,w} \equiv K^0_{d,w}$ be the open-string
Gromov-Witten invariant of Eq. \!\! (\ref{virloc})
in genus zero at degree $d$ with winding $w > 0.$
Recall that $K_{d,w}$ has only a calculational
definition.  Using this definition, it is simple to
show that
\begin{equation}
\label{gwdef}
K_{d,w} = {\mathcal C}_w \int_{\overline{\mathcal M}_{0,1}
({\mathbb P}^2,d)} e(E_d) \cdot
\frac{ev^*(\phi_p)}{\lambda-\psi}
\end{equation}
where $e(E_d)$ is the equivariant Euler class of
the obstruction bundle $E_d \equiv
R^1\rho_*ev^*K_{{\mathbb P}^2};$
$\phi_p$ is the equivariant class of the
north pole;
$ev^*$ is the pull-back under the evaluation
map $ev:  {\overline{\mathcal M}_{0,1}
({\mathbb P}^2,d)} \rightarrow \PP^2;$
and $$\lambda \equiv -\lambda_z/w$$
is the weight of the tangent space of the
leg at the point of attachment.  $\mathcal{C}_w$
is the overall factor discussed above, and is
defined by
\begin{equation}
\label{Cw}
{\mathcal C}_{w} \equiv \frac{1}{(-\la_n)(-\la_z)}
\cdot
C_w,\quad {\rm where}
\quad C_w \equiv \frac{\la_n + \la_z}{-w\la_z}
\prod_{k=1}^{w-1}\frac{\frac{k}{w}\la_z + \la_n}{\frac{k}{w}\la_z}
\end{equation}
(the factor of $w$ in the denominator accounts
for automorphisms of the leg).
In the above,
$ \int_{\overline{\mathcal M}_{0,1}
({\mathbb P}^2,d)}$
represents the equivariant push-forward under
the evaluation map, and
will henceforth be written $ev_{*}.$
Noting that $i_p^*(\phi_p)$ has the effect of
multiplying by $(-\lambda_z)(-\lambda_n),$
we can cancel out these terms and write
\begin{equation}
\label{onepoint}K_{d,w} =
C_w \cdot i_p^*ev_{*}\left(
\frac{e(E_d)}{\lambda-\psi}\right).
\end{equation}
Note that here we are interpreting $K_{d,w}$ as an
element of $\Q (\lambda_1,\lambda_2,\lambda_3)$ since
our prescription for computing it yields not a number,
but a rational function of the weights of the torus.
Similarly, the other side of this equation is
an element of the localized equivariant cohomology
of a point, so the above equation makes sense.

\subsection{Equivariant Mirror Theorem}

The mirror theorems of Givental \cite{G1} \cite{G2}
and Lian-Liu-Yau \cite{LLY} are actually proven in 
the equivariant setting.
Using their results we have an algebraic procedure
for calculating $ev_{*}\left(
\frac{e(E_d)}{\lambda-\psi}\right),$
hence $K_{d,w},$
which we now describe.  (The theorem does not apply at
higher genus, where the explicit graph sum is the only
approach currently available.)

Let $H$ be the equivariant hyperplane class on
$\PP^2$ (so $H$ restricted to the $i$th fixed point
is $\lambda_i$).  Let $K$ be the equivariant first
Chern class of the canonical bundle with
its canonical torus action.  Let us define $J$
and $\widetilde{J}$ by
$$
J\equiv e^{\frac{t_0 + H t_1}{x}} \widetilde{J}
\equiv e^{\frac{t_0 + H t_1}{x}}  (1 + K \cdot \sum_{d>0} q^d 
ev_*\frac{e(E_d)}{x(x-\psi)})
$$
as well as $I$ and $\widetilde{I}$ by
\begin{equation}
\label{Jeq}
I \equiv e^{\frac{t_0 + H t_1}{x}} \widetilde{I}
\equiv e^\frac{t_0 + H t_1}{x} 
\sum_{d\geq 0} q^d \frac{\prod_{m=0}^{3d-1}(K - mx)}
{\prod_{m=1}^{d} \prod_{i=1}^{3} (H-\lambda_i +mx)},
\end{equation}
where $q = e^{t_1}.$
Finally, define $I_1$ by
\begin{equation}
\label{Ieq}
I=  e^\frac{t_0 + H t_1}{x}(1+I_1 \frac{H}{x} + o(\frac{1}{x^2})).
\end{equation}
Explicity, 
$$
I_1=\frac{K}{H} \sum_{d=1}^\infty \frac{(-1)^{d+1}(3d-1)!}{(d!)^3}q^d
$$
Then the equivariant mirror theorem states that
\begin{equation}
\label{equivmirthm}
J(t_0, t_1 + I_1) = I(t_0,t_1).
\end{equation}
(See \cite{E} or \cite{LLY} for this formula, but note that in
their calculations, a different linearization on $K_{\PP^2}$
is used.)
In implementing this change of variables,
one must not neglect that $q=e^{t_1}$
should be replaced by $e^{t_1+I_1} = qe^{I_1(q)}$
on the left side.
To make things more concrete, we can substitute
$\lambda = -\lambda_z/w$ for $x$ and replace
equivariant classes by their restrictions via $i_p^*.$
Then $H = \lambda_3,$ $K = \lambda_1 + \lambda_2
-2\lambda_3,$ and as always $\lambda_z = \lambda_1-
\lambda_3$ and $\lambda_n = \lambda_2 - \lambda_3.$
Now defining $Q(q)$ by $$Q(q) = qe^{I_1(q)},$$
one can write Eq. \!(\ref{equivmirthm})
as $J(Q) = I(q).$  In practice, it is useful
to invert and solve for $q(Q)$ (see Eq.\!({\ref{zofQ}}))
to the order
needed in calculations.

This procedure yields the same $K_{d,w}$ as the
graph sum.

\subsection{Mirror Symmetry}

Now that we have a power-series prescription for computing
the open-string Gromov-Witten invariants -- defined by the
graph sum and expressible via Eq. \!\!(\ref{onepoint})
in terms of equivariant invariants of
one-pointed maps -- we can ask if this procedure yields
the same numbers that we get from physics
via the B-model and mirror symmetry \cite{AV} \cite{AKV}.

Let us review the physics approach.  
First form the generating function
\begin{equation}
\label{gwgenfun}
W(Q,y) = \sum_{d,w} K_{d,w} Q^d y^w,
\end{equation}
where $Q = e^{-\hat{t}}$ is
the K\"ahler parameter
and $y = e^{\hat{u}}$ is the complexified
holonomy parameter.
%
We have
\begin{equation}
W = \int \hat{v}(\hat{u}) d\hat{u},\qquad {\rm or}\qquad
\partial_{\hat{u}} W = \hat{v}(\hat{u}),
\end{equation}
where
$\hat{v}$ and $\hat{u}$
are {\bf classically} equal to the
un-hatted parameters
obeying the following equation:
\begin{equation}
\label{cpxeq}
e^u + e^v + 1 + e^{-t - u - v} = 0;
\end{equation}
or
\begin{equation}
\label{cpxeqsol}
v =
\log\left[-(1+e^u)/2 - \frac{1}{2}
\sqrt{(1+e^u)^2-4e^{-t}e^{-u}}\right].
\end{equation}
[The origin of the equation is a specialization
of $X_1 + X_2 + X_3 + X_4 = 0$ in the patch
$X_4 = 1$ subject to the constraint
$X_1 X_2 X_3 X_4^{-3} = e^{-t}.$
Here we have written $X_1 = e^u, X_2 = e^v, X_3 = e^{-t-u-v}.$
Note that the exponents of this constraint, $(1,1,1,-3),$
are precisely the toric data for $K_{\PP^2}$.]

In fact, the relation of the parameters in
Eq. \!\!(\ref{cpxeq}) and their hatted cousins
receives instanton corrections.
More precisely, $u$ and $v,$ the parameters in the
complex equation, are not the
``flat coordinates.''
The map from the parameters in the equation to the
flat K\"ahler parameters is what we need to perform
enumerative tests.
We have
$\hat{u}=\hat{u}(u,t)$ and $\hat{v}=\hat{v}(v,t).$
The relation between the K\"ahler
parameter $\hat{t}$ and the complex parameter $z = e^{-t}$
is well known:  $\hat{t}$ is the logarithmic
solution to the Picard-Fuchs equation for the Fermat
cubic in ${\mathbb P}^2$).  The other relations
were found in \cite{AKV}.
Explicitly,
for our computation, we have
\begin{eqnarray}
\label{hateq1}
\hat{u}(u,t) &=& u + (t - \hat{t})/3 + i\pi \qquad
{\rm or}\qquad e^{\hat{u}} = -e^u e^{\Delta/3},\\
\label{hateq2}
\hat{v}(v,t) &=& v + (t - \hat{t})/3 + i\pi  \qquad
{\rm or}\qquad e^{\hat{v}} = -e^v e^{\Delta/3},\\
\label{hateq3}
\Delta &\equiv& t - \hat{t} =  \sum_{k=1}^{\infty} = \frac{(-1)^k}{k}
\frac{(3k)!}{(k!)^3}z^k,
\end{eqnarray}
where again $z = e^{-t}.$
The Picard-Fuchs equation is
${\mathcal L}f = 0,$ where ${\mathcal L} = \theta^3 
+ 3z(3\theta + 1)(3\theta + 2)$ and $\theta \equiv
z\frac{d}{dz}.$

Now if we use Eq. \!(\ref{cpxeqsol}) with
Eqs. (\ref{hateq1})-(\ref{hateq3}), we can
solve for $\hat{v}$ in terms of $\hat{u}.$
First, 
following \cite{AKV},
we define $r \equiv e^{\Delta/3} =
(\frac{Q}{z})^{\frac{1}{3}} = 1 - 2Q + 5Q^2 - 32Q^3 + ...,$
where we have used the inverse relation
\begin{equation}
\label{zofQ}
z(Q) = Q + 6Q^2 + 9Q^3 + 56Q^4 - ...
\end{equation}
($Q(z)$ is obtaind from Eq. \!(\ref{hateq3}) by exponentiation). 
We get
\begin{equation}
\label{solveforvhat}
\hat{v} = \log\left[(r-y)/2 + \sqrt{(r-y)^2/4 + Q/y}\right].
\end{equation}
Then Eq. \!(\ref{gwgenfun})
and $\hat{v} = \partial_{\hat{u}}W$
tell us
\begin{equation}
\label{hatveq}
\hat{v} = \sum_{d,w}w K_{d,w}Q^d y^w.
\end{equation}
From Eqs. (\ref{solveforvhat})
and (\ref{hatveq}) we can read off the Gromov-Witten
invariants, $K_{d,w}.$\footnote{To
connect with the notation of the previous section,
set $t = -t_1,$ so $z = q.$  Thus the inverse relation
$z(Q)$ given in Eq. \!(\ref{zofQ})
is the same one needed for the equivariant mirror theorem.}

\subsection{Equivalence}

We now show that the two methods of computing
open-string invariants are the same.  We are grateful
to A. Klemm for helping us with this verification.

To make this explicit, we will use a specific set
of weights, namely $\lambda_z = 1,$ $\lambda_n = 0$
(this corresponds to ``ambiguity zero'' in the physics).
In this case, 
$K = 1$ and $H = 0,$ as explained below
Eq. \!(\ref{equivmirthm}),
while $C_w = -1/w.$

Note
$\widetilde{J}(Q,x=-1/w) = 1 + \sum_{d>0} Q^d w^2 K_{d,w}.$
Here it is convenient to note $K_{0,w} = 1/w^2,$
which is the pure leg contribution.
Then
$$\sum_{w\neq 0} \widetilde{J}(Q,-1/w)y^w =
\sum_{d \geq 0; w \neq 0} w^2 K_{d,w}.$$
Comparing with Eq. \!(\ref{hatveq}), we see that this is
$\partial_{\hat{u}}{\hat{v}},$ or
equivalently $\partial^2_{\hat{u}} W(Q,y)$
(recall $y = e^{\hat{u}}$).
Now let's take Eq. \!(\ref{solveforvhat}) and
find $\partial_{\hat{u}} \hat{v}.$
Some algebra leads to 
$$\partial_{\hat{u}} \hat{v} =
-\frac{1}{2} - \frac{r-3y}{4\sqrt{(r-y)^2/4 + Q/y}},$$

On the other hand, the mirror theorem of
Eq. \!(\ref{equivmirthm}) together with
Eq. \!(\ref{Ieq}) with our choice of weights
gives
$$\sum_{w\neq 0} 
\widetilde{J}(Q,-1/w)y^w=
\sum_{w\neq 0,d\geq 0} y^w r^{-w} q^d (-1)^d w
\frac{(w+3d-1)!}{(d!)^2(w+d)!}.$$

Thus we predict that, up to constants, we have the
following equivalence of power series:\footnote{Equality
holds if we add $1/2$ to the left hand side}
$$-\frac{r-3y}{4\sqrt{(r-y)^2/4 + Q/y}}=
\sum_{w\neq 0\atop d\geq 0}
y^w r^{-w} q^d (-1)^d w
\frac{(w+3d-1)!}{(d!)^2(w+d)!},
$$
with $q(Q)$ and $r(Q)$ given by Eq. \!(\ref{zofQ})
and $r = (Q/q)^{1/3}.$
To prove this, we define $a=y/r$ and note that the
left hand side is equal to
$$\frac{3a-1}{2(1-a)}\left(1+\frac{4q}{a(1-a)^2}\right)^{-1/2}
= \frac{3a-1}{2}\sum_{d=0}^{\infty}\frac{(2d-1)!!}{(2d)!!}
(-1)^d\frac{4^d\,q^d}{a^d(1-a)^{2d+1}},$$
where we have used $(1+x)^{-1/2}=\sum_{n=0}^\infty
\frac{(2n-1)!!}{(2n)!!}(-1)^nx^n,$ with $(2n)!!\equiv
2^n(n!).$  Now expanding $(1-a)^{-k} = \sum_{n=0}^\infty
\frac{(k+n-1)!}{n!(k-1)!}a^n,$
the coefficient of $a^{m-d}q^d$ is seen to be
$$\frac{(-4)^d(2d-1)!!}{2\cdot (2d)!!}
\left[\frac{3(2d+m-1)!}{(2d)!(m-1)!}-
\frac{(2d+m)!}{(2d)!m!}\right]=
(-1)^d(m-d)\frac{(m+2d-1)!}{(d!)^2 m!}.$$
The same coefficient appears on the right hand side,
with $w=m-d.$
Curiously, the explicit form of $Q(q)$ is not used.

For nonzero ambiguity $s$, one makes the
change of $y \rightarrow ye^{s\hat{v}}$
in Eq. (\ref{solveforvhat}) and re-solves for $\hat v$
before performing the check, where now $C_w$ of Eq. (\ref{Cw})
is taken with $\lambda_z=1,$ $\lambda_n=s$. 
\section{Calculations}

We have implemented a computer program running in Maple
which fully automates the calculation described
in Section \ref{weights}.
A similar program for graphs without legs was
described in \cite{KZ}.  The program
computes open-string Gromov-Witten invariants,
as listed in Table 1.

Let us first display some calculations in ``ambiguity zero.''
The ambiguity $p$ encountered in physical calculations
corresponds to $\lambda_z = 1$ and  $\lambda_n = p.$
(When $\lambda_z \neq 1$ the calculations do not lead to
integer invariants.) 

\vskip 0.2in

{\vbox{
$$
\vbox{\offinterlineskip\tabskip=0pt
\halign{\strut
\vrule#&
&\hfil ~$#$
&\hfil ~$#$
&\hfil ~$#$
&\vrule#\cr
\noalign{\hrule}
&g
&d=1
&d=2
&
\cr
\noalign{\hrule}
&0
&-1
&\frac{1}{4}(w^2 + 4w + 15)
&\cr
&1
&\frac{1}{24}(w^2 - 2)
&-\frac{1}{96}(3w^4 + 20w^3 + 53w^2 + 24w - 24)
&\cr
&2
&-\frac{1}{5760}(3w^4-20w^2+24)
&\frac{w}{23040}(39w^5+364w^4+1185w^3+1200w^2-632w-1056)
&\cr
\noalign{\hrule}}\hrule}$$}
\centerline{{\bf Table 1:} Some
open-string Gromov-Witten invariants $K^g_{d,w}$} 
\centerline{for general (positive)
winding, $w$
(with ambiguity $p=0$).}
\vskip7pt}

A surprising result of these calculations
is the weight dependence of the results.  This is in
contrast to the usual, non-equivariant
calculations of Gromov-Witten
invariants.
Physically, this ambiguity is related to different possible
``special coordinates'' in the mirror B-model, and to
the framing of a link in a Chern-Simons theory 
\cite{AKV}
(the Chern-Simons calculation
is only relevant to the ${\cal O}_{\PP^1}(-1,-1)$ example
of \cite{KL} \cite{LS}).

At genus zero, our results agree with the predictions
of Aganagic, Klemm, and Vafa, including
the dependence on the torus weights,
i.e. on the ``ambiguity'' $p = \lambda_n.$
Tables appear in their paper \cite{AKV}.

At degree zero, for the genera computed ($g\leq 4$),
our results
agree with the predictions of Ooguri and Vafa \cite{OV}
and the computations of Katz and Liu \cite{KL}
and Li and Song \cite{LS}.
Namely, at ambiguity zero, degree zero,
winding $w$ and genus $g$ the Gromov-Witten
invariant $K^{g}_{0,w}(p=0)$
is given as the negative
of the $u^{2g-2+1}$ term in the expansion
of $(1/w)[2\sin (wu/2)]^{-1}.$
For example, $K^4_{0,3} = -3429/5734400.$

At higher genera and degree, there are no known predictions from
physics of what the higher genus
open-string Gromov-Witten
formulas should be, though the work of Ooguri and
Vafa \cite{OV}
and of Labastida, Mari\~no and Vafa \cite{LMV}
leads to strong integrality checks, which hold
for the invariants we compute.
Let $n^{g}_{d,w}$ denote the integer invariants
representing the number of BPS domain walls.
To demonstrate integrality,
consider the invariants $K^g_{2,3},$
i.e. the right column of the Table 1, with $w=3.$
Since $d$ and $w$
are relatively prime, there are only contributions
to this term from $n^{g}_{d,w}$ with
$(d,w) = (2,3).$  Using the integrality relations
of \cite{LMV} (Eq. (2.11) of \cite{MV}),
we find from Table 1,
$K^0_{2,3} = - n^0_{2,3} = 9,$ so $n^0_{2,3} = -9.$
Then
$K^1_{2,3} = (7/24) n^{0}_{2,3} + n^1_{2,3} = -109/8,$
so $n^1_{2,3} = -11.$  Finally,
$K^2_{2,3} = (-29/1920)n^0_{2,3} + (-3/8)n^{1}_{2,3}
- n^2_{2,3} = 6567/640,$ and we find
$n^2_{2,3} = -6.$  If instead we consider $w=2,$ then
there are more contributions to the Gromov-Witten
invariant due to multiple windings.
The integer invariants are listed in Table 2.

\vskip 0.2in

{\vbox{
$$
\vbox{\offinterlineskip\tabskip=0pt
\halign{\strut
\vrule#&
&\hfil ~$#$
&\hfil ~$#$
&\hfil ~$#$
&\vrule#\cr
\noalign{\hrule}
&g
&d=1
&d=2
&
\cr
\noalign{\hrule}
&0
&1
&-\frac{1}{4}(w^2 + 4w + 16 - \epsilon_{w})
&\cr
&1
&0
&-\frac{1}{48}(w^4 + 8w^3 + 20w^2 + 16w + 3 \epsilon_{w})
&\cr
&2
&0
&-\frac{1}{2880}(2w^6+24w^5+95w^4+120w^3-52w^2-144w -
45 \epsilon_{w})
&\cr
\noalign{\hrule}}\hrule}$$}
\centerline{{\bf Table 2:} Some
integer invariants $n^g_{d,w}$ for various general (positive)} 
\centerline{winding, $w$
(with ambiguity $p=0$). Here $\epsilon_{w}=
(1-(-1)^{w})/2.$}
\vskip7pt}

It is also easy to compute the
open-string Gromov-Witten
invariants corresponding to multiple boundary components
(``holes'').  Once again, the integrality tests have
held up in each calculation checked.
For example, we find for a surface with two boundary
components with windings $\vec{n}=(1,w)$ that
$n^{g=1}_{d=2,\vec{n}=(1,w)} =$
$(-1/24)(w^4+10w^3+35w^2+50w+48),$ an integer.

Of course, it would be extremely desirable to develop
recursive
differential equations for topological partition functions
which would be an open-string version of the BCOV equations
\cite{BCOV}.
The lack of (2,2) worldsheet symmetry will complicate
matters significantly.
Perhaps the calculations of this paper can
play some role in the establishing
such equations,\footnote{Two recent papers have
taken steps towards doings so \cite{GJS}
\cite{M}} or in
resolving an open-string version
of the holomorphic ambiguity
as analogous calculations
do in the closed string case \cite{KZ}.

\vskip0.2in
\centerline{\bf Acknowledgments}
\vskip0.1in
Thanks to M. Aganagic, A. Klemm, and C. Vafa for
many discussions of their work; to E. Getzler; to J. Bryan;
and to R. Pandharipande
and A. Klemm for their help with Section \ref{mirsec}.
The work of E.Z.
is supported in part by NSF Grant DMS-0072504 and
an Alfred P. Sloan Foundation Fellowship. T.G. is
partially supported by an NSF postdoctoral fellowship.

\end{document}